# Magnetic Imaging with High Spatial Resolution Using a Single Niobium-Nitride-Based Three-Dimensional Nanobridge Junction

Yinping Pan[1,2], Yue Wang[1,2], Xinxin Fan[1,2,3], Xu Liu[1,2,3], Shirui Liu[1,2,3], Xiaoyu Liu[1,2], Ling Wu[1,2], Lu Zhang[1,2,3,*], Wei Peng[1,2,3], Jie Ren[1,2,3], Jiamin Xiong[1,2], Hao Li[1,2,3], Lixing You[1,2,3], Zhen Wang[1,2,3,4], and Lei Chen[1,2,3,*]

1 State Key Laboratory of Materials for Integrated Circuits, Shanghai Institute of Microsystem and Information Technology, Chinese Academy of Sciences, Shanghai 200050, China

2 Shanghai Key Laboratory of Superconductor Integrated Circuit Technology, Shanghai Institute of Microsystem and Information Technology, Chinese Academy of Sciences, Shanghai 200050, China

3 University of the Chinese Academy of Sciences, Beijing 100049, China

4 School of Physical Science and Technology, Shanghai Tech University, Shanghai 200031, China

*Corresponding authors: zhanglu@mail.sim.ac.cn; leichen@mail.sim.ac.cn

**Abstract**

The nanobridge junctions (NBJs) allow the miniaturization of the Superconducting Quantum Interference Device (SQUID) and improve the spatial resolution of scanning SQUID microscopy. However, the SQUID loop diameter imposed a fundamental limit on the resolution. In this work, we proposed a single niobium-nitride-based three-dimensional nanobridge junction (NbN-3D-NBJ) alternative to the nanoSQUID for high-resolution magnetic imaging. The critical current of an NbN-3D-NBJ showed an nearly monotonic dependence on the external magnetic field, enabling nanoscale mapping of the magnetic field distribution by measuring the junction's critical current with an intrinsic noise level of 100 nT/√Hz. Using a NbN-3D-NBJ probe, we successfully imaged an Abrikosov vortex in a niobium film with a minimal full width at half maxima (*FWHM*) of 0.53 μm, constrained by the distance from the junction to the probe tip edge. Furthermore, we resolved 80 nm spacings between the superconducting nanowires utilizing a second probe with an optimized junction-to-tip distance. Thus, the single 3D NbN NBJ probe has been demonstrated as an effective approach for magnetic imaging with ultra-high spatial resolution.

## Introduction

The scanning SQUID microscope (sSM) has gained rising attention as a unique tool for magnetic imaging at cryogenic temperatures, empowering the investigation of superconductivity, topological phenomena, and magnetic materials[1-3]. By measuring local magnetic fields with extremely high sensitivity, sSMs have provided critical insights, including confirmation of the d-wave pairing symmetry in YBCO[4], discovery of the coexistence of superconductivity and ferromagnetism[5], observation of temperature-dependent fraction of the flux quantum[6], and visualization of electronic transport in a quantum anomalous Hall insulator[7], etc. However, due to fabrication challenges, the spatial resolution of a conventional on-chip SQUID probe is limited to the micrometer scale[8-9]. Recent advancements in the SQUID-on-tip (SOT) probe have improved spatial resolution by integrating a nanoSQUID onto the apex of a quartz pipette[10-11]. Therefore, SOT-based imaging has mapped Landau levels and magnetic oscillation in magic-angle graphene[12-14], revealing the orbital ferromagnetism in moirè Chern insulator[15] and hydrodynamic flow of electrons in type II Weyl semimetal[16], etc.

The nanoSQUID at the apex of SOT is composed of two parallel nano-bridge junctions significantly reducing the size of the SQUID loop and enhancing sensitivity[17-19]. Despite advancements in SOT, further optimization of the SQUID loop to diameters below 100 nm presents significant challenges. An alternative approach involves using single Josephson junctions, which exhibit magnetic field response characteristics due to flux quantization within the junction[20-21]. Compared to SQUIDs, scaling the dimensions of a single junction is more straightforward, as it avoids the complexities of maintaining constant separation between two junctions in a loop. Regarding spin sensitivity, single junctions outperform traditional SQUID loops, as their near-field coupling coefficients are significantly higher[22-24]. A notable demonstration involved scanning an MFM (magnetic force microscopy) tip over a planar niobium-copper-niobium junction, illustrating its potential for high-resolution magnetic imaging[25]. The achieved spatial resolution was still constrained by the dimension of the planar junction. The potential resolution can also be enhanced by the ferromagnetic layer in superconductor ferromagnet/superconductor junctions by focusing magnetic flux[26], but it also introduces complications such as magnetic interaction with sample characterization.

This study employs a niobium-nitride-based three-dimensional nanobridge junction (NbN-3D-NBJ) as a scanning probe for nanoscale magnetic imaging. The NbN-3D-NBJ exhibits a nearly monotonic relationship between its critical current and external magnetic fields, enabling direct spatial mapping of magnetic field distributions by monitoring the junction's critical current variations. We imaged Abrikosov vortices in the niobium film with a minimal full width at half maxima of 520 nm, which was proven to be limited primarily by the junction-to-tip distance. Furthermore, we used a probe with an optimized junction-to-tip distance and resolved the 80 nm spacing between superconducting nanowires. With improved operational simplicity and enhanced spatial resolution, the single NbN-3D-NBJ probe offers a high-resolution magnetic imaging technique for investigating quantum materials and phenomena.

## Results

Figure 1a displays a top-view scanning electron microscopy (SEM) image of a single NbN-3D-NBJ probe utilized in our experiments. The NbN-3D-NBJ structure was fabricated in tight proximity to the physical vertex of the probe tip, with a measured junction-to-tip distance $r$ of 950 nm from the nanobridge junction center to the vertex edge. The NBJ exhibits a width $w$ of 121 nm. Figure 1b illustrates the cross-sectional geometry of the 3D NBJ, including the junction thickness $t$ = 18 nm and length $l$ = 17.5 nm. The bank thicknesses of the 3D NBJ were measured to be 150 nm and 100 nm for the respective sides, which are much greater than the thickness of the bridge deck. The 3D structure of the NBJ improved the non-linearity of its current-phase relation[27-29].

As schematically represented in Figure 1c, the NbN 3D NBJ probe was mounted with a roughly 15-degree inclination relative to the sample surface. The vertical position of the tip vertex above the sample surface $h_t$ was precisely controlled by monitoring the resonance amplitude of a tuning fork (TF). This configuration establishes the junction-to-sample distance ($h_J = r \times \sin 15^{\circ} + h_t$) as a crucial parameter governing the spatial resolution of magnetic imaging. More details regarding the fabrication process and experimental setup for the 3D NBJs are provided in the Methods section.

The electrical transport properties of the NbN 3D NBJ were characterized by measuring the current-voltage characteristics $I_b$-$V_{NBJ}$ under sweeping perpendicular magnetic fields $B_z$. The relatively high critical field of NbN material enables the 3D NBJ made of NbN to maintain the superconducting state under applied magnetic fields up to 1 T. Figure 1d presents the contour map of $V_{NBJ}$ as a function of both $I_b$ and $B_z$. The yellow area with low voltage values indicates the superconducting state of the NBJ. The red and blue areas represent the normal state of the NBJ with a positive and negative bias current, respectively. It showed that the magnetic field dependence of the junction critical current $I_c$ can be roughly fitted by the dashed curves in Figure 1d, which is calculated by the Fraunhofer diffraction pattern of a single Josephson junction in the field [27]: $I_c(B_z) - I_0 = (I_{c\text{-max}} - I_0)\times |\sin(\pi B_z/B_0)/(\pi B_z/B_0)|$, where $I_{c\text{-max}}$ = 35.5 μA represents the maximum critical current, and $B_0$ = 0.32 T denotes the period of the magnetic field corresponding to a flux quantum[30]. In weak-link planar junctions, the magnetic field dependence of critical current ($I_c$) transitions from a Fraunhofer-diffraction-like pattern to a monotonic decay one as the junction width decreases[31]. In Figure 1d, the significantly diminished amplitude of the side peaks suggests that the NBJ dimension is approaching this crossover regime. Furthermore, the $I_c(B_z)$ of the NBJ is asymmetric about zero magnetic field, probably due to the screening effect from the asymmetric current distribution in its superconducting banks[32-33]. At a fixed bias current of $I_b$ = 35 μA, $V_{NBJ}$ exhibits an asymmetric dependence on $B_z$ with derivative $\partial V_{NBJ}/\partial B_z$ = 33 mV/T at zero field (inset of Figure 1e), making it suitable for magnetometry applications. To assess the noise level of a single NBJ as a magnetometer, we measured its voltage noise at zero magnetic fields (Figure 1e). Using a room-temperature preamplifier with an input noise of 2 nV/Hz$^{1/2}$, the voltage noise across the NBJ was recorded as 3.3 nV/Hz$^{1/2}$ in a magnetically shielded room and 80 nV/Hz$^{1/2}$ in the scanning system. The corresponding magnetic noise levels were 100 nT/Hz$^{1/2}$ and 2.4 μT/Hz$^{1/2}$, respectively. The extrinsic noise is currently about one order of magnitude higher than the intrinsic noise of the NBJ tip. In the scanning system, the room-temperature preamplifier is located far from the tip, and environmental radio-frequency noise couples into the wiring before the preamplifier. This limitation can be mitigated by using a SQUID array amplifier installed at low temperature near the NBJ tip.

By scanning a single NBJ magnetometer, we obtained a magnetic image of an Abrikosov vortex in a 150-nm-thick niobium (Nb) film at a bath temperature of 2 K. The image scan covered an area of 2×2 μm², with each picture requiring 22 minutes to acquire. We captured a series of images while gradually reducing the tip-to-sample distance $h_t$ from 151 nm to 39 nm, as shown in Figure 2a. Below 39 nm, the tip began to scratch the sample surface, preventing the acquisition of clear images. As $h_t$ decreased, both the image intensity and spatial resolution of the Abrikosov vortex's magnetic image improved. Figure 2b presents line cuts through the vortex center along the x and y directions, shown in blue and red, respectively. These magnetic profiles were well-fitted using Gaussian peak functions. The extracted full width at half maximum (*FWHM*) and peak height of $\Delta B_z$ were plotted in Figure 2c as functions of $h_t$. According to Figure 1, the junction-to-sample height for the NBJ probe used in this image is given by $h_J \approx$ 246 nm + $h_t$. Consequently, the NBJ probe was positioned approximately 285 nm above the sample surface, achieving magnetic imaging of an Abrikosov vortex with a minimum *FWHM* of 0.53 μm. The observed linear dependence of *FWHM* on $h_t$ suggests that the current spatial resolution of magnetic imaging is primarily limited by the junction-to-tip distance $r$ of the probe. Our result is consistent with the simulation work reported in Ref. [33]. Additionally, the *FWHM*s of the magnetic profiles in the x-direction were found to be 160 nm larger than those in the y-direction. This anisotropic distortion of the vortex image is likely attributed to differences in the width $w$ and length $l$ of the NBJ.[33]

Next, we performed the image scanning over the Nb film under an applied magnetic field $B_z$. Before each image, the Nb film was heated to 15 K using a heating resistor installed beneath it and cooled to the bath temperature under $B_z$. Figure 3a presents the vortex distribution pattern in the Nb film under $B_z$ ranging from 0 to 7mT. During the imaging experiment, the tip-to-sample distance $h_t$ was consistently maintained at approximately 150 nm. The image scan covered an area of 5×5 μm², with each picture requiring 10 minutes to acquire. As $B_z$ increased, the number of vortex-pinning sites in the Nb film also rose. At higher magnetic fields, some vortices formed clusters to minimize the system's total energy. Interestingly, the diameter of individual vortices appeared to decrease with the increase of the magnetic field. To understand this effect, we performed a line-cut analysis of the magnetic profiles along the dashed lines and plotted the results as blue dots in Figure 3b. At $B_z$ = 0 mT, the *FWHM* of a single vortex was measured to be 0.82 μm, with a Gaussian fit shown as red curves. At $B_z$ = 1 mT, the magnetic profile of two adjacent vortices was well-fitted by overlapping Gaussian functions with the same *FWHM*. As $B_z$ increased to 2 mT and 3 mT, the overlap between vortices became more pronounced while still following the profiles of the same Gaussian functions. With a further increase in $B_z$, the adjacent vortices moved closer, causing their magnetic signals to merge, making only the top section of the vortices visible. By $B_z$ = 5 mT and 7 mT, the spacing between adjacent distinguishable vortices reduced to 0.57 μm and 0.49 μm, respectively, which is smaller than the *FWHM* of an individual vortex.

As shown in Figure 4a, we utilized the NbN nanowire pattern from a superconducting nanowire single-photon detector as a benchmark to calibrate the minimal spatial resolution of a single NbN 3D NBJ probe[34-35]. To enhance the spatial resolution, we employed a different probe with a shorter junction-to-tip distance of

r = 320 nm, as depicted in Figure 4b. Consequently, with $h_J$ = 83 nm + $h_t$, the junction were allowed to approach closer to the sample surface. The field dependence of the critical current exhibited behavior to the previous probe in Figure 1. Figure 4c presents an SEM image of the NbN nanowires, which have a width of 200 nm and are uniformly spaced 80 nm apart. We applied a magnetic field of $B_z$=-10 mT to probe the diamagnetic response of the superconducting NbN nanowires.

The magnetic image of Figure 4d, acquired over a 1×1.1 μm$^2$ scan area in 18 minutes, reveals a clear nanowire profile closely resembling the SEM image. The 80 nm spacing between adjacent NbN nanowires is distinctly resolved, demonstrating that the probe's spatial resolution is better than 80 nm. The 80 nm spacings between adjacent NbN nanowires are distinctly resolved, demonstrating that the probe's spatial resolution is better than 80 nm. This capability makes it highly promising for high-resolution magnetic imaging of superconducting devices and for investigating vortex dynamics in these structures.

**Discuss**

The ability to resolve the 80 nm spacing between NbN nanowires in our imaging experiments highlights the feasibility and advantage of the NbN-based 3D NBJ probe for high-resolution magnetic imaging, demonstrating its potential as a powerful tool for nanoscale magnetic characterization. The fabrication process of 3D NBJs is scalable and compatible with batch production. The nearly monotonic dependence of the junction's critical current on the applied magnetic field allows for absolute magnetic field measurement through a simple voltage readout. The use of NbN as the superconducting material plays a crucial role in the performance of the NBJ. The high critical field of NbN ensures that the superconducting state of the NBJ remains stable under perpendicular magnetic fields exceeding 1 T.[36] Given that the magnetic field required for a single flux quantum to enter the NBJ is approximately 0.32 T, the ability of NbN 3D NBJs to keep superconductivity under high magnetic fields is essential for reliable magnetic imaging.

The experimentally confirmed 80 nm resolution already exceeds that of most on-chip SQUID probes (typically around 500 nm) and is comparable to the state-of-the-art SQUID-on-tip devices (~50 nm)[1]. By further reducing the size of the nanobridge junction, the spatial resolution can be readily improved. In comparison with scanning NV-center magnetometers, the NbN NBJ tip can operate at cryogenic temperatures below 9 K and under out-of-plane magnetic fields up to 0.32 T, offering distinct advantages for characterizing quantum materials under extreme conditions.

During the vortex imaging experiments, we observed an asymmetric distortion of the vortex profiles along the x and y directions. Since we used an optical interferometer to monitor the sample position precisely, this suggests that the distortion is related to the NBJ sensor rather than an artifact of position miscalibration. It is probably due to the anisotropy of the NBJ structure, where differences in the junction's length and width may introduce variations in the coupling coefficients along different directions[37]. Additionally, flux-focusing effects caused by the large bank areas of the 3D NBJ could also contribute to this asymmetry. Further investigations, including modeling and simulations of the coupling coefficients and local magnetic field distributions, are necessary to understand this effect fully.

Several optimizations can be implemented to further enhance spatial resolution in future experiments. Reducing the junction-to-tip distance *r* and optimizing the probe's mounting angle will position the junction closer to the sample, thereby improving resolution. Another key approach is minimizing the junction size, which will not only improve its current-phase relation but also strengthen its coupling with magnetic dipoles[38]. Theoretically, as the junction size decreases, the field dependence of the critical current will transition to a monotonic decay regime, improving the field-to-voltage transfer function and further enhancing measurement sensitivity. However, whether a minimal critical size exists in the magnetic response of the NBJ remains an open question, and its presence would impose a fundamental limit on the spatial resolution. The magnetic response of an individual NBJ originates from the trapping of flux quanta within the junction, with its intrinsic magnetic field noise theoretically comparable with that of a SQUID, primarily limited by the thermal noise of the junction's normal resistance[19, 39]. However, in our scanning microscopy setup, system noise remains the dominant limiting factor, arising from mechanical vibrations of the scanning stage, electrical noise in the NBJ readout wiring, and radio-frequency interference. Effectively suppressing these noise sources will be crucial for further enhancing the performance of magnetic imaging. Furthermore, by scanning a magnetic structure with a known field profile, the in-situ calibration and control of hj can be realized, which in turn enables the further optimization of spatial resolution in magnetic imaging[40].

## Conclusion

The NbN-based three-dimensional (3D) nanobridge junction (NBJ) probe exhibits a spatial resolution better than 80 nm, demonstrating exceptional potential for ultra-high-resolution magnetic imaging at the nanoscale. Its unique 3D architecture enhances the nonlinearity of the junction's current-phase relationship, leading to a Fraunhofer diffraction pattern in the critical current as a function of the applied magnetic field, which enables direct measurement of absolute magnetic fields by monitoring the voltage across the NBJ, with an intrinsic noise level of 100 $nT/Hz^{1/2}$. Furthermore, magnetic imaging of Abrikosov vortices as a function of tip-to-sample distance revealed a minimum full width at half maximum (FWHM) of 520 nm, primarily constrained by the probe's junction-to-tip distance *r*. By employing a second probe with an optimized junction-to-tip distance *r* and bringing it closer to the sample surface, the NBJ probe successfully resolved the 80 nm spacing between superconducting nanowires, highlighting its capability for high-resolution imaging. Future advancements, such as reducing junction dimensions, optimizing probe mounting angle, and mitigating system noise, will be essential to further improve spatial resolution and sensitivity. Therefore, the NbN 3D NBJ probe represents a powerful and practical tool for next-generation nanoscale magnetometry and magnetic imaging[40].

**Methods**

The fabrication process of the NbN 3D NBJ probe followed the same procedure as the Nb 3D nano-SQUID[28-29]. It begins with the deposition of a 150 nm NbN layer on a silicon substrate coated with 300 nm of $SiO_2$. Next, photolithography (PL) and reactive ion etching (RIE) are used to pattern the first NbN wiring layer, while the photoresist is intentionally retained. A 25 nm $SiO_2$ insulating barrier is then deposited to cover the resist, followed by the deposition of a second 150 nm NbN layer, which is subsequently patterned using PL and RIE. The lift-off process is then applied to remove the previous photoresist, creating an insulating gap for the NBJ as shown in Figure 1b. Afterward, the NbN NBJ layer is deposited and patterned using electron-beam lithography (EBL). The probe tip shape is then defined through PL, and RIE is employed to remove the 300 nm $SiO_2$ coating. Finally, deep silicon etching is performed through the silicon wafer, completing the fabrication of probe devices.

Figure 5b shows the schematic of the scanning system for a single NbN-3D-NBJ probe; it is the same setup as a scanning SQUID microscope in the Refs[41-42]. The NBJ tip is mounted onto one prong of a tuning fork (TF), while the fine motions of a piezo-stack move the samples. The TF resonance amplitude and frequency are monitored when the probe tip approaches the sample surface. The resonance of TF is turned off during the imaging scanning. The probe and sample were both thermally anchored to a copper (Cu) base with a 2 K bath temperature.

The single NBJ was current-biased by connecting a 20 kΩ resistor in series with a voltage source, ensuring a stable current supply. The voltage across the NBJ was measured using a low-noise preamplifier with a 100× gain and subsequently digitized by an analog-to-digital converter. Meanwhile, the sample's position during scanning was precisely tracked using an optical interferometric position sensor integrated into a scanning probe microscopy controller (Attocube ASC 500). By synchronizing the recorded voltage signals with the positional data, the magnetic image was finally reconstructed with accurate spatial dimensions.

**Author Contributions**

Y. P. and L. C. performed the experiments and collected the data. L. Z. and L. C. planned the research and wrote the paper. Y. W., X. F., X. L., and S. L. participated in and assisted with the experiments. X. L., L. W., and W. P. assisted in the fabrication of devices. J. X., H. L., and L.Y. provide the SNSPD sample. J. R. and Z. W. assisted in the data analysis. All authors approved the final version of the manuscript.

**Acknowledgments**

This work is supported by the National Natural Science Foundation of China (Grant No. 62301542) and, Shanghai Postdoctoral Excellence Program (No. 2022678). L. C. acknowledges the support by the National Natural Science Foundation of China (Grant Nos. 62071458 and 11827805). L. Z. acknowledges the support by the Natural Science Foundation of Shanghai (Grant No. 22ZR1473400), and the Young Investigator program of the CAS (Grant No. 2022235). L. Y. acknowledges the support by the Innovation Program for Quantum Science and Technology (No. 2023ZD0300100). The device fabrication was performed in the Superconducting Electronics Facility (SELF) of SIMIT.

**Additional Information**

Competing Interests: The authors declare no competing interests

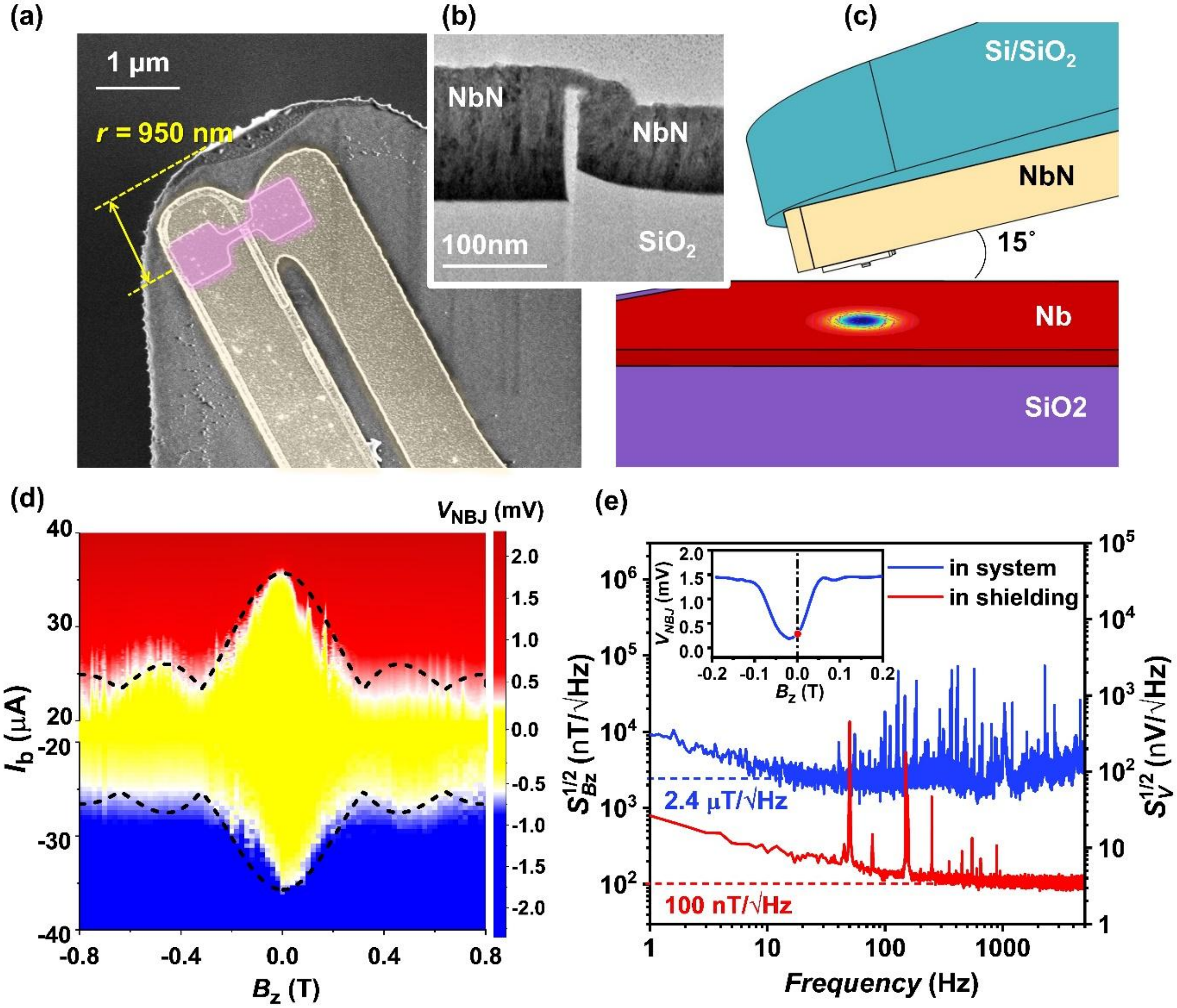

Figure 1. NbN-based 3D NBJ probe. (a) SEM image of the NBJ probe's top view. (b) Cross-sectional view of the NBJ, showing NbN bank electrodes at both ends, separated by a $SiO_2$ insulating layer, with the NBJ deck positioned on top. (c) Schematic of the NBJ probe setup (not-in-scale) for imaging vortices on a niobium film, with a tilt angle of approximately 15°. (d) Contour plot of $V_{NBJ}$ ($I_b$, $B_z$) characteristics, with dashed lines representing Fraunhofer diffraction fitting curves of $I_c(B_z)$. (e) Magnetic field noise density spectrum in a magnetically shielded room and the scanning system. The inset shows the $V_{NBJ}$-$B_z$ curve at a fixed $I_b$ of 35 μA.

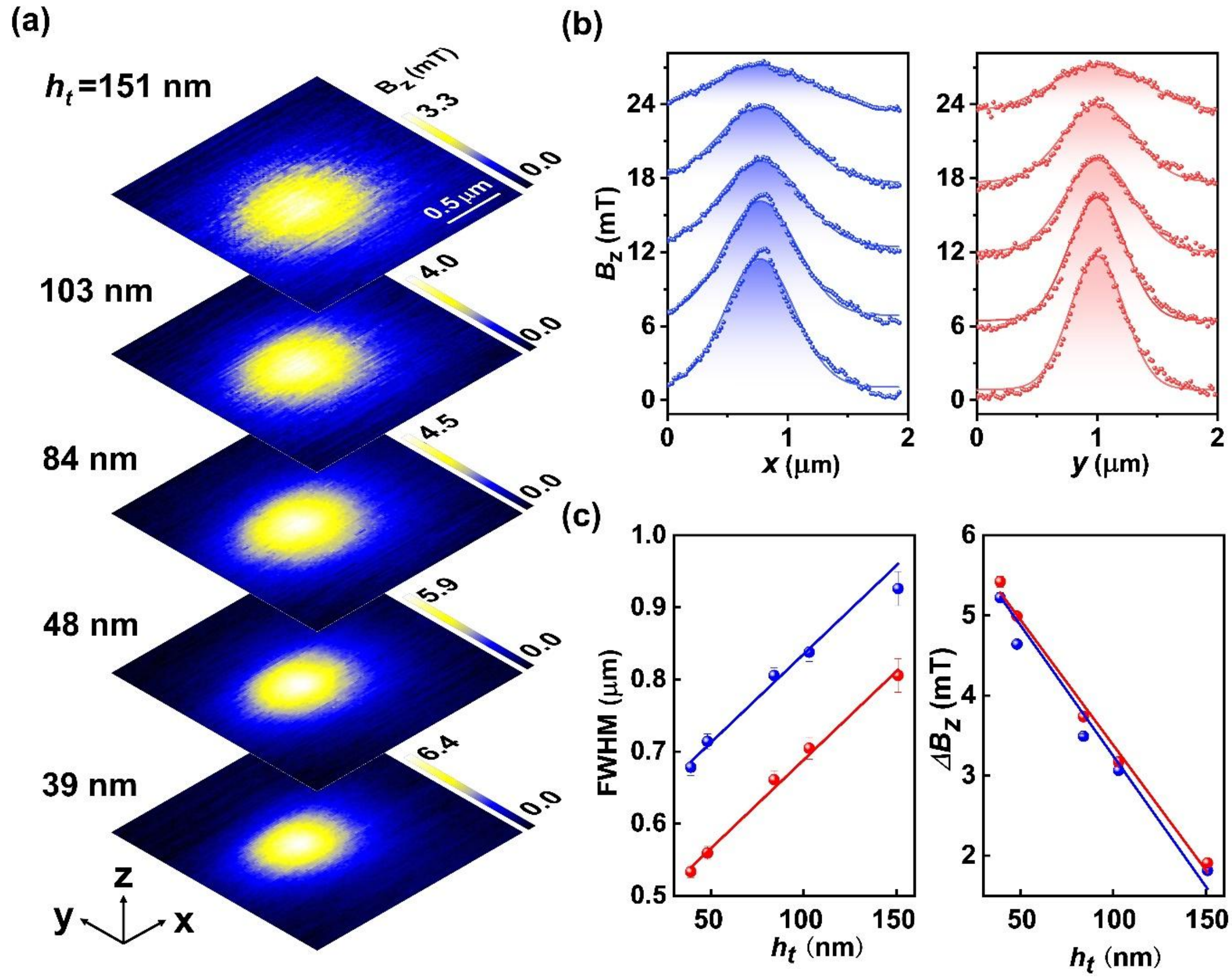

Figure 2. Magnetic imaging of an Abrikosov vortex. (a) Vortex images at various tip-to-sample distances along the *z*-axis, were obtained by scanning the NBJ probe along the x-axis with the nanobridge oriented parallel to the y-axis. (b) Line-cut profiles extracted from (a) through the vortex center along the *x* and *y* directions, fitted with Gaussian functions. The curves are vertically offset for clarity. (c) The full width at half maximum (*FWHM*) and amplitude ($\Delta B_z$) of the Gaussian fits in (b) are plotted as functions of the tip-to-sample distance $h_t$. The dependence of FWHM and $\Delta B_z$ on $h_t$ are fitted with straight lines.

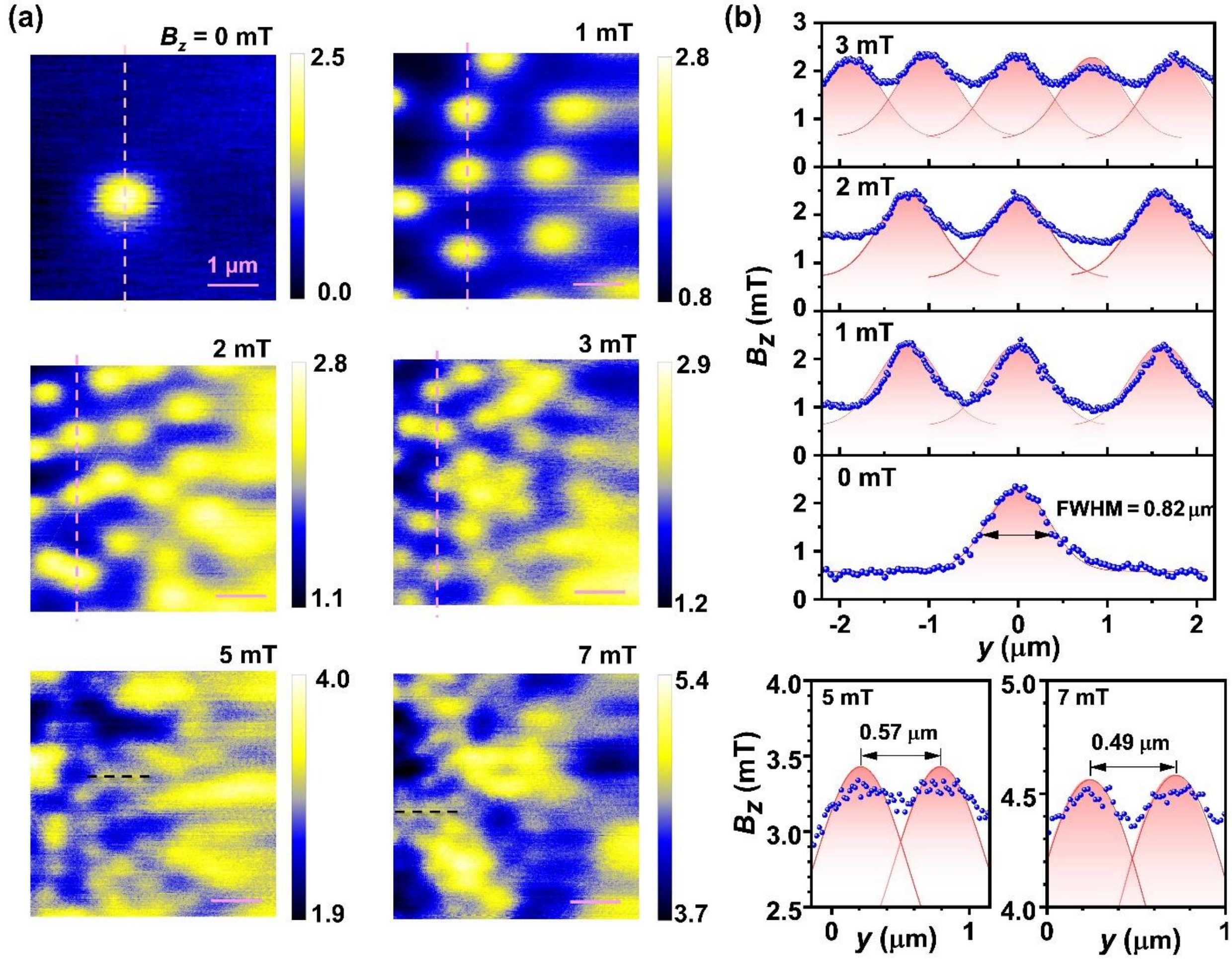

Figure 3: Magnetic imaging of Abrikosov vortices under various magnetic fields. (a) The images of vortex distribution in the Nb film were taken at a constant $h_t$=150 nm. Before each imaging, the Nb film was heated to 15 K and cooled to 2 K at the corresponding field. (b) Blue dots represent line-cut profiles extracted from the dashed lines in (a) at different $B_z$ values. The red curves correspond to the same Gaussian fit obtained for $B_z$= 0 mT, with an *FWHM* of 0.82 μm. As $B_z$ increases, the profiles of adjacent vortices begin to overlap, and the image can distinguish the adjacent vortices of distances less than the *FWHM*.

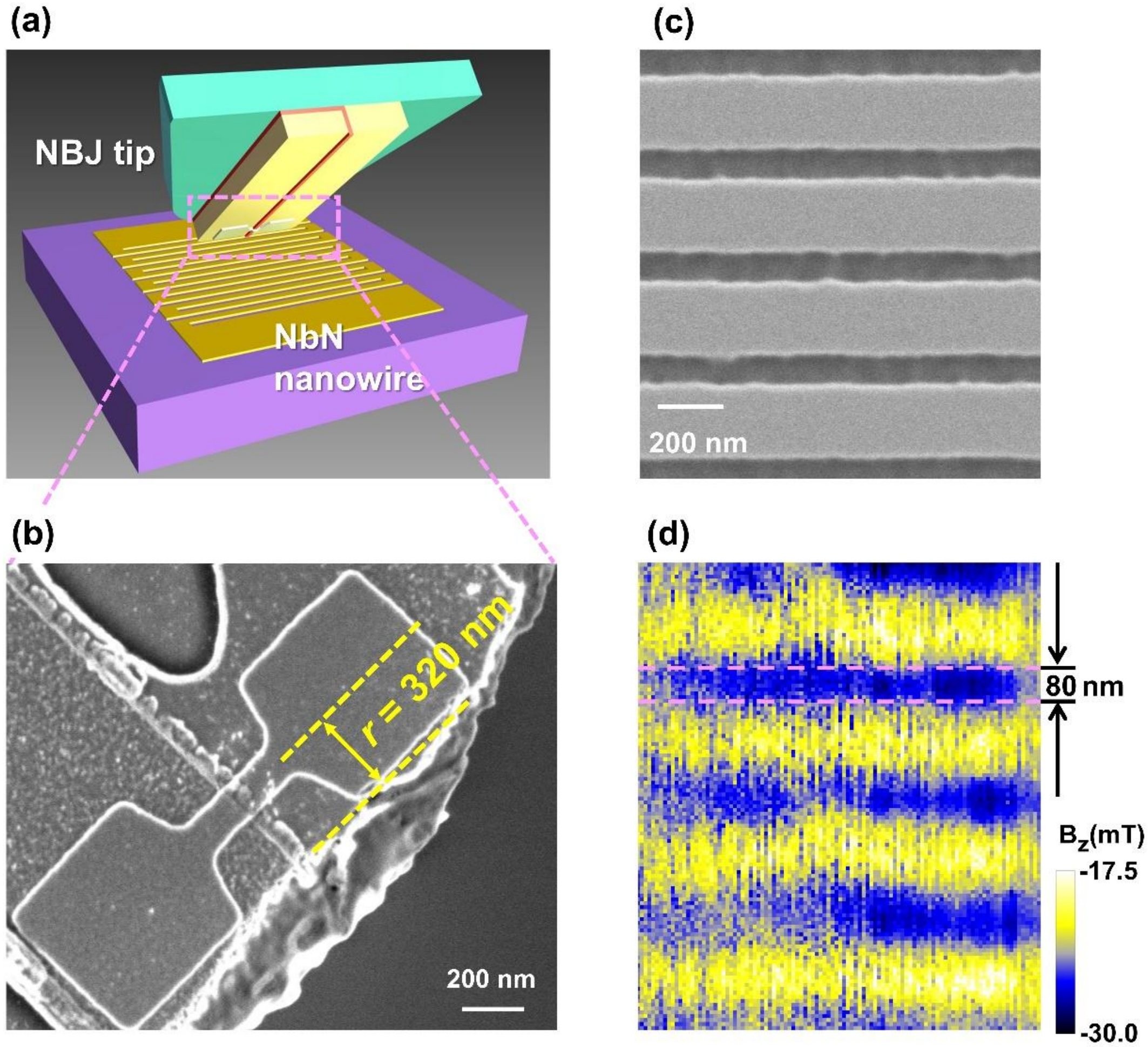

Figure 4: Magnetic imaging of superconducting nanowires. (a) Schematic illustration of the measurement setup (not to scale) (b) SEM top-view of the second NBJ probe with an optimized tip-to-junction *r* = 320 nm. (c) SEM image of superconducting nanowires, featuring a width of 200 nm and an inter-wire spacing of 80 nm. (d) Magnetic imaging of the diamagnetic response from superconducting nanowires at $B_z$ = -10 mT, clearly resolving the 80 nm spacing highlighted in blue.

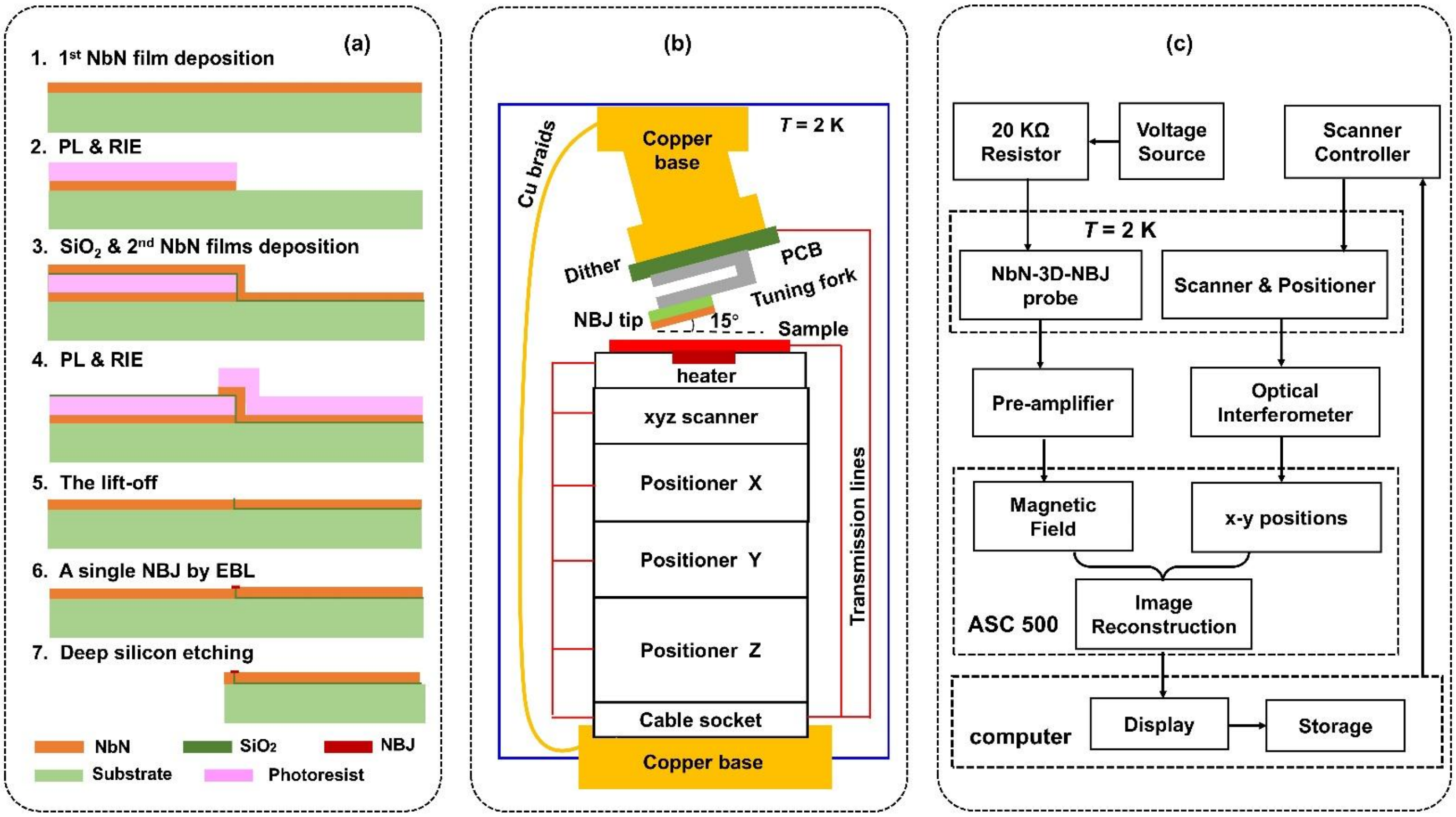

Figure 5: (a) Main steps in the fabrication process of the NbN-3D-NBJ probe. (b)Schematic of the scanning microscope housing at 2 K. (c) Diagram of the readout and controller system for the NBJ probe and imaging scanner.

**Table of content**

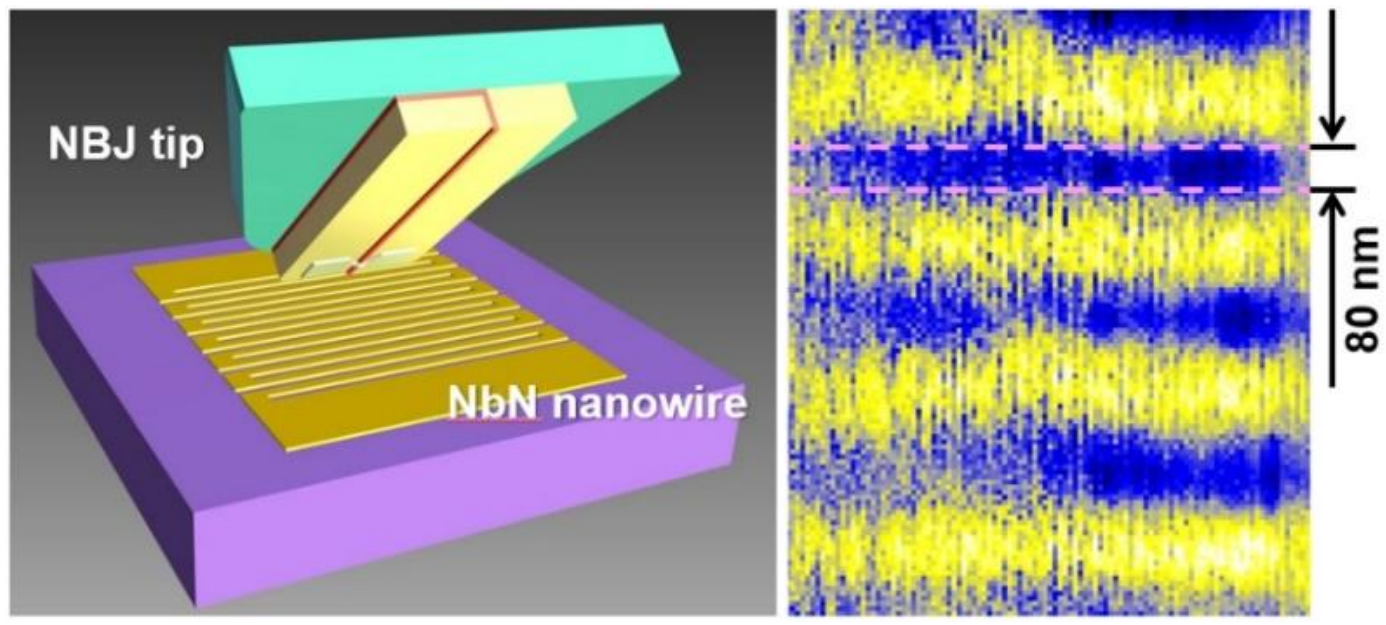